\documentclass[%
preprint%
 ,secnumarabic%
,superscriptaddress%
,amssymb, amsmath,nofootinbib,aps]{revtex4}

\newcommand{\be}{\begin{equation}}
\newcommand{\ee}{\end{equation}}

 \newcommand{\bea}{\begin{eqnarray}}
\newcommand{\eea}{\end{eqnarray}}

\usepackage{epsfig}%
\usepackage{graphicx}%
\usepackage{color}
\usepackage[colorlinks=true,linkcolor=blue]{hyperref}%
\expandafter\ifx\csname package@font\endcsname\relax\else
 \expandafter\expandafter
 \expandafter\usepackage
 \expandafter\expandafter
 \expandafter{\csname package@font\endcsname}%
\fi

\begin{document}

\title{A minimal model for ${\rm SU}(N)$ vector dark matter}
\author{Stefano Di Chiara}
\email{stefano.dichiara@helsinki.fi}
\affiliation{Helsinki Institute of Physics, P.O.Box 64, FI-00014, University of Helsinki, Finland}
\affiliation{Department of Physics, University of Jyv\"askyl\"a, P.O. Box 35, FI-40014, University of Jyv\"askyl\"a, Finland}
\author{Kimmo Tuominen}
\email{kimmo.i.tuominen@helsinki.fi}
\affiliation{Department of Physics, University of Helsinki, P.O. Box 64, FI-000140, University of Helsinki, Finland}
\affiliation{Helsinki Institute of Physics, P.O.Box 64, FI-00014, University of Helsinki, Finland}

\def\be{\begin{equation}}
\def\ee{\end{equation}}
\def\bea{\begin{eqnarray}}
\def\eea{\end{eqnarray}}


\begin{abstract}
We study an extension of the Standard Model featuring a hidden sector that consists of a new scalar charged under a new SU$(N)_D$ gauge group , singlet under all Standard Model gauge interactions, and coupled with the Standard Model only via a Higgs portal.  We assume that the theory is classically conformal, with electroweak symmetry breaking dynamically induced via the Coleman-Weinberg mechanism operating in the hidden sector. Due to the symmetry breaking pattern, the SU$(N)_D$ gauge group is completely Higgsed and the resulting massive vectors of the hidden sector constitute a stable dark matter candidate. We perform a thorough scan over the parameter space of the model at different values of $N=2$, $3$, and $4$, and investigate the phenomenological constraints. We find that $N=2,3$ provide the most appealing model setting in light of present data from colliders and dark matter direct search experiments. We expect a heavy Higgs to be discovered at LHC by the end of Run II or the $N=3$ model to be ruled out.
\end{abstract}
\maketitle

\section{Introduction}
In addition to the Higgs boson \cite{ATLAS,CMS}, LHC has so far not discovered any signals for new physics at the terascale. This result has recently led to explore novel possible solutions to the naturalness problem \cite{Bardeen:1995kv,Foot:2007iy,Farina:2013mla,Heikinheimo:2013fta}. The essential assumptions of these approaches are the absence of physical mass scales above the electroweak (EW) scale and that the boundary conditions at the Planck scale lead to the vanishing of the quadratic divergence to the Higgs boson mass.

Within a classically conformal theory, one sets all explicit mass terms to zero in the tree level Lagrangian. One must then address the question of how the weak scale arises. One possibility is that the weak scale is generated radiatively \cite{Coleman:1973jx}, but this does not work quantitatively for the Standard Model (SM). On the other hand, motivated by the lack of the SM to explain the observed dark matter abundance or matter-antimatter asymmetry, one may introduce additional sectors very weakly coupled with the SM. Maintaining the classical conformality also in the hidden sector, one can then generate a nontrivial scale radiatively and this is transmitted to the SM sector via interactions between the two sectors \cite{Englert:2013gz}. This is the mechanism which we consider in this paper.

More concretely, we extend the SM by a hidden sector consisting of a scalar transforming nontrivially under a new non-abelian gauge symmetry. All SM fields are singlet under this new gauge symmetry, and the radiatively generated vacuum expectation value of the hidden sector scalar leads to a complete breaking of the hidden gauge symmetry. The resulting massive gauge bosons are mass degenerate and due to a residual global symmetry, they constitute a dark matter candidate \cite{Hambye:2008bq}. We set up the theory for general hidden gauge group SU$(N)_D$, extending earlier work \cite{Hambye:2013dgv,Carone:2013wla} where the $N=2$ case was considered. We then investigate the phenomenological viability of the model numerically for $N=2$, $3$, and $4$ by imposing the stability of the potential up to the Planck scale, requiring perturbativity of all couplings, and imposing the constraints from the LHC data. Furthermore we compute the dark matter relic density and impose constraints from the presently known abundance \cite{Ade:2015xua}, as well as from the direct searches for dark matter \cite{Aprile:2012nq,Amole:2015lsj,Akerib:2013tjd}.

The paper is organised as follows: the model and computation of the EW symmetry breaking as well as the dark matter relic density are presented in Sec. \ref{model}. Various phenomenological constraints are considered in Sec. \ref{UFBD}, and in Sec. \ref{checkout} we present the conclusions and outlook.


\section{SU$(N)$ vector dark matter}\label{model}

\subsection{Preliminary}

Before the EW gauging, the global symmetry of the SM scalar sector is SU(2)$_L\times$SU(2)$_R$, which can be made explicit by assembling the Higgs fields into a matrix $H$ transforming as a bifundamental of this symmetry. We now generalise this as follows: Consider extending the matter content of the SM by a scalar fields assembled into a matrix $\Phi$, singlet under the SM gauge group and transforming as a bi-adjoint under the  global SU(N)$_L\times$SU(N)$_R$ symmetry.  Then we gauge the SU(N)$_L$ symmetry and denote this new gauge group by SU$(N)_D$. Explicitly, we then have
\be\label{phi}
\Phi=\frac{\sigma}{\sqrt{N^2-1}} I+i \frac{\phi_a}{\sqrt{2 N}} T^a\ ,\quad \Phi^\prime= \exp\left[-i g_D\alpha_a T^a \right]\Phi\ ,\quad a=1,\ldots,N^2-1
\ee 
with real fields $\sigma$ and $\phi_a$, $I$ the identity matrix in $N^2-1$ dimensions, $T^a$ a generator of the adjoint representation\footnote{Note that this holds only for real or pseudoreal representations. If we consider $\Phi$ transforming as bi-fundamental under the global symmetry, we would need to use complex fields $\sigma$ and $\phi^a$ in Eq.~\eqref{phi} for $N>2$. This would double the real degrees of freedom and since we have a minimal model in mind, we do not pursue this further.}
of  SU$(N)_D$, and their numerical factors chosen to preserve canonical normalization for any $N$. On the other hand we assume the SM matter fields to be singlets under SU$(N)_D$, so that the Lagrangian includes all the SM kinetic, gauge and Yukawa terms, together with 
\be\label{LP}
{\cal L}\supset {\rm Tr}\left[D_\mu\Phi\right]^\dagger D^\mu\Phi-V\ ,\quad D^\mu=\partial^\mu-ig_D A^\mu_aT^a\ .
\ee
While mass terms for both scalar fields are allowed, we set them to zero at tree level to make the model classically conformal:
\be\label{Vt}
V=\frac{\lambda_h}{2} \left(H^\dagger H\right)^2+\frac{\lambda_\phi}{2}{\rm Tr}\left(\Phi^\dagger\Phi\right)^2-\lambda_p H^\dagger H {\rm Tr}\Phi^\dagger\Phi\ ,\quad H=\frac{1}{\sqrt{2}}\left(\begin{array}{c}  \pi^+ \\ h+i\pi^0 \end{array} \right) \ .
\ee
The last term of the potential, generally referred to as the portal coupling \cite{Silveira:1985rk,Patt:2006fw,Barger:2007im,Englert:2013gz}, generates mass terms for $H$ and $\Phi$, once both scalars develop a vacuum expectation value (vev). 

\subsection{Electroweak symmetry breaking}

While the tree level potential in Eq.~\eqref{Vt} has its minimum at the origin of field space, the scalar vevs acquire non-zero values via dimensional transmutation because of quantum corrections \cite{Coleman:1973jx}. The one loop contribution to the effective potential in the $\overline{\rm MS}$ scheme \cite{Martin:2001vx} can be written as
\be
\Delta V=\sum_{p\in\{\varphi,\psi,A\}} (-1)^{2 s_p}\frac{2 s_p+1}{64\pi^2}m_p^4\left(\log\frac{m_p^2}{\Lambda^2}-k_p\right)\ ,\quad k_\varphi=k_\psi=\frac{3}{2}\ ,\ k_A=\frac{5}{2}\ ,
\ee
where the sum over $p$ includes scalars ($\varphi$), fermions ($\psi$), and vectors ($A$). The factor $s_p$ denotes the spin of the particle in question, and $m_p$ its field dependent tree-level mass. The resulting one loop effective potential, 
\be
V_{1L}=V+\Delta V\ ,
\ee
reaches a minimum at
\be\label{vevs}
\langle H \rangle=\frac{1}{\sqrt{2}}\left(\begin{array}{c}  0 \\ v_h \end{array} \right)\ ,\quad \langle \Phi \rangle=\frac{v_\phi}{\sqrt{N^2-1}} I\ ,
\ee
provided that the values of the vevs, assumed to be real, satisfy the minimization conditions for the tree level potential,
\be\label{minc}
\left.\frac{\partial V}{\partial \varphi_i}\right|_{vev}=0\ ;\ \varphi_i=h,\sigma \quad \Rightarrow\quad \lambda _{\phi }=\frac{v_w^2}{v_{\phi }^2}\lambda _p\ ,\ \lambda _h=\frac{v_{\phi }^2}{v_w^2}\lambda _p\ .
\ee
The scalar mass matrix at the minimum of the potential is then defined as (with no sum over indices)
\be\label{Ms}
\left({\cal M}^2_\varphi\right)_{i j}=\left.\frac{\partial^2 V_{1L}}{\partial \varphi_i \partial\varphi_j}\right|_{vev}-\left.\frac{\delta_{i j}}{v_i}\frac{\partial \Delta V}{\partial \varphi_i}\right|_{vev}\ ,
\ee
where the last term represents the shift generated by the one loop correction on the otherwise zero tree level mass terms \cite{Elliott:1993bs}. 

The vevs in Eq.~(\ref{vevs}) ensure the breaking of the SM gauge group following the usual pattern, and of SU$(N)_D$ entirely. Consequently, all the dark gauge bosons $A^a$ acquire the same mass
\be\label{mA}
m_A=\frac{g_D v_\phi}{\sqrt{N-N^{-1}}}\ .
\ee
This degeneracy is a consequence of the residual SO$(N)$ global symmetry of the Lagrangian, which guarantees the stability of the SU$(N)_D$ gauge boson multiplet. These massive gauge bosons are therefore suitable dark matter candidates \cite{Hambye:2008bq}.

The pseudoscalars $\phi_a$ provide the longitudinal degree of freedom to $A_a$, while $\pi^0$ and $\pi^\pm$ are absorbed by EW gauge bosons $Z$ and $W^\pm$, respectively.
In Appendix~\ref{mass} we provide the analytical result for the one loop scalar mass matrix, Eq.~\eqref{Ms}, in the $(h,\sigma)$ basis. From Eqs.~(\ref{scalarmasses}) the one loop masses and the corresponding mass eigenstates, $h_1$ and $h_2$, can be easily derived analytically.

From the results above we see that viable EW symmetry breaking is possible without the intervention of any mass term at the tree level. 
Under renormalisation then, the higher order corrections to the scalar masses depend on the renormalization scale only logarithmically, and therefore are in principle natural \cite{Englert:2013gz,Heikinheimo:2013fta}. In this sense classical conformality trades the SM fine tuning problem with finding justification for taking the mass terms equal to zero to begin with.

\subsection{Dark matter abundance}\label{DMs}
As we pointed out in the previous subsection, the residual SO$(N)$ makes the massive $A^a$ vector bosons stable and therefore suitable dark matter candidates. Their annihilation and semi-annihilation cross sections can be easily calculated in the limit of zero portal coupling $\lambda_p$. This approximation for the dark matter analysis is consistent, as in the next section it turns out that $\lambda_p\ll g_D$ in the viable region of parameter space. 

For the thermally averaged annihilation ($AA\rightarrow \sigma\sigma$) and semi-annihilation ($AA\rightarrow \sigma A$) cross section times relative velocity we find
\be
\langle \sigma v\rangle_{\rm{ann}} =\frac{11 m_A^2}{144 \left(N^2-1\right) \pi  v_{\phi }^4}\ ,\quad \langle \sigma v\rangle_{\rm{semi-ann}} =\frac{3 m_A^2}{8 \left(N^2-1\right) \pi  v_{\phi }^4}\ ,
\ee
which for $N=2$ reproduce the results of \cite{Hambye:2013dgv,Carone:2013wla}. This approximation is sufficient when working away from the resonance thresholds, where the full thermal average \cite{Gondolo:1990dk} should be used; see e.g. \cite{Cline:2013gha}.

The dark matter abundance is determined by 
\be
\frac{dY}{dx}=Z(x)\left(\langle \sigma v\rangle_{\rm{ann}} (Y^2-Y_{\rm{eq}}^2)+\frac{1}{2}\langle \sigma v\rangle_{\rm{semi-ann}} Y(Y-Y_{\rm{eq}})\rangle\right),
\ee
where $Y=n/s$, $Y_{\rm{eq}}$ the corresponding equilibrium density, $x=m_A/T$, and 
\be
Z(x)=-\sqrt{\frac{\pi}{45}}M_{\rm{Pl}} m_A\sqrt{g_\ast(m_A/x)}x^{-2} \ ,
\ee
with $g_*$ denoting the effective number of degrees of freedom and $M_{Pl}$ the Planck mass. 

Using the standard approximations, the dark matter abundance is determined by
\be\label{Ohs}
\Omega h^2=N\frac{1.07\times 10^9 {\rm GeV}^{-1} x_f}{\sqrt{g_*\left(x_f\right)}M_{\rm{Pl}}\langle v\sigma \rangle}\ ,\quad \langle \sigma v\rangle=\langle \sigma v \rangle_{\rm{ann}}+\frac{1}{2}
\langle \sigma v\rangle_{\rm{semi-ann}} \ .
\ee
The value of $x_f=m_A/T_f$ is determined by solving 
 \be
 x_f=\ln\left[\frac{Z(x_f)y_{\rm{eq}}(x_f)^2}{y^\prime_{\rm{eq}}(x_f)-y_{\rm{eq}}(x_f)}\left(\frac{\delta(\delta+2)}{\delta+1}\langle \sigma v\rangle_{\rm{ann}}+\frac{\delta}{2}\langle \sigma v\rangle_{\rm{semi-ann}}\right)\right],
 \ee
where $\delta$ determines the deviation of the distribution from the equilibrium one, $\delta=Y/Y_{\rm{eq}}-1$, before the freeze out. The value of $\delta$ is expected to be ${\cal{O}}(1)$, and in the numerical analysis we choose $\delta=1$.


\section{Constraints}\label{UFBD}
\subsection{The LHC data fit}
To perform the quantitative analysis of the viability of the model, we start by scanning the parameter space for data points producing a viable mass spectrum. Between the two possible hierarchy choices, we focus on the case when the Higgs scalar is heavier than the dark scalar; we comment on the viability of the alternative possiblity in the next section. 
Given the tight experimental constraints on the masses of the SM particles, we set all the SM couplings (except $\lambda_h$) as well as the Higgs vev to their standard values ($v_h=246$ GeV). We fix $\lambda_h$ and $\lambda_\phi$ via Eqs.~\eqref{minc} and $v_\phi$ by setting $m_{h_1}=125$ GeV. This leaves us with only two free parameters: $g_D$ and $\lambda_p$. We then collect $10^5$ data points for each value of $N=2$, $3$, $4$ in the region
\be
0<g_D<1.4,\,\,\, 0<\lambda_p<0.12\ .
\ee
We will see in the next subsection that the scanned range of values of $g_D$ and $\lambda_p$ is sufficient to cover the phenomenologically viable region.

The off-diagonal terms in the scalar mass matrix, Eq.~\eqref{Ms1L}, are proportional to the portal coupling, $\lambda_p$, which therefore controls the amount of mixing between the SM Higgs field $h$ and the dark scalar $\sigma$ in the mass eigenstate $h_1$, parametrized by the angle $\alpha$ according to
\be
\left(
\begin{array}{c}
 h_1 \\
 h_2
\end{array}
\right)=\left(
\begin{array}{cc}
 \cos\alpha & -\sin\alpha \\
 \sin\alpha  & \cos\alpha
\end{array}
\right) \left(
\begin{array}{c}
 h \\
 \sigma 
\end{array}
\right) .
\ee
 Given that $\sigma$ does not couple to SM particles, the physical Higgs $h_1$ couplings turn out to be suppressed as compared to their SM values by a factor of $\cos\alpha$. We constrain this factor, and consequently $\lambda_p$, by determining for each data point the goodness of fit of the Higgs coupling strengths to their corresponding measured values for the $\gamma\gamma$, $ZZ$, $WW$, $bb$, $\tau\tau$ inclusive processes \cite{ATLAS,CMS,Tevatron}. 
To calculate $\chi^2$, we follow the procedure detailed in \cite{Alanne:2013dra}, and here we present directly the results of the statistical analysis. Among the $10^5$ scanned data points about 39\%, 40\%, 39\%, for $N=2,3,4$, respectively, satisfy the 95\%CL constraint
\be\label{LHCc}
p\left(\chi^2>\chi^2_j\right)>0.05\ ,\quad 1\leqslant j\leqslant 10^5,
\ee
with the index labeling the $j$-th data point for a given $N$. The average values of the mixing coefficient, portal coupling, dark gauge coupling, and scalar vev among the viable data points are
\be\label{LHCfit}
N=\left\{\begin{array}{c}
2\\ 3\\ 4
\end{array}\right.
\ , \quad 
\overline{\cos\alpha}=\begin{array}{c}
0.95\\ 0.95\\ 0.95
\end{array}\ ,\quad 
\overline{\lambda_p}=\begin{array}{c}
0.063\\ 0.064\\ 0.059
\end{array}\ ,\quad 
\overline{g_D}=\begin{array}{c}
0.58\\ 0.64\\ 0.66
\end{array}\ ,\quad 
\overline{v_\phi}/{\rm GeV}=\begin{array}{c}
1335\\ 1310\\ 1328
\end{array} .
\ee
As expected, given that the measured Higgs couplings are SM like, the portal coupling is constrained by collider data to acquire small values. The quartic coupling $\lambda_\phi$ turns out to be even smaller than $\lambda_p$, from Eqs.~\eqref{minc}, given that the dark vev $v_\phi$ is much larger than $v_h$.

In the next section we further constrain the viable data points by requiring perturbativity of all the couplings and stability of the potential up to the Planck scale.


\subsection{Stabilization of the SM potential}\label{ViaPS}
\begin{figure}[htb]
\centering
\includegraphics[width=0.46\textwidth]{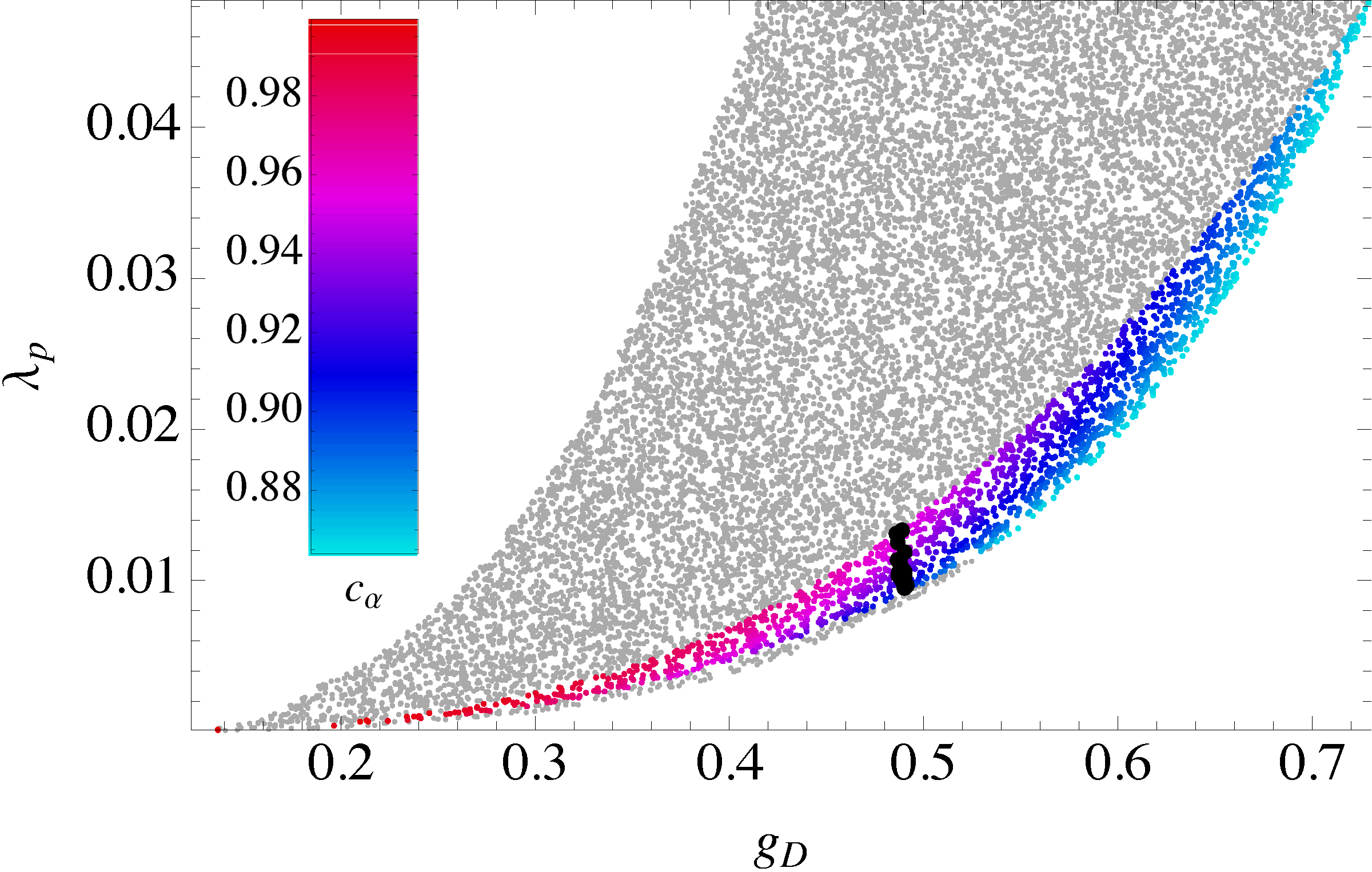}\hspace{1cm}
\includegraphics[width=0.46\textwidth]{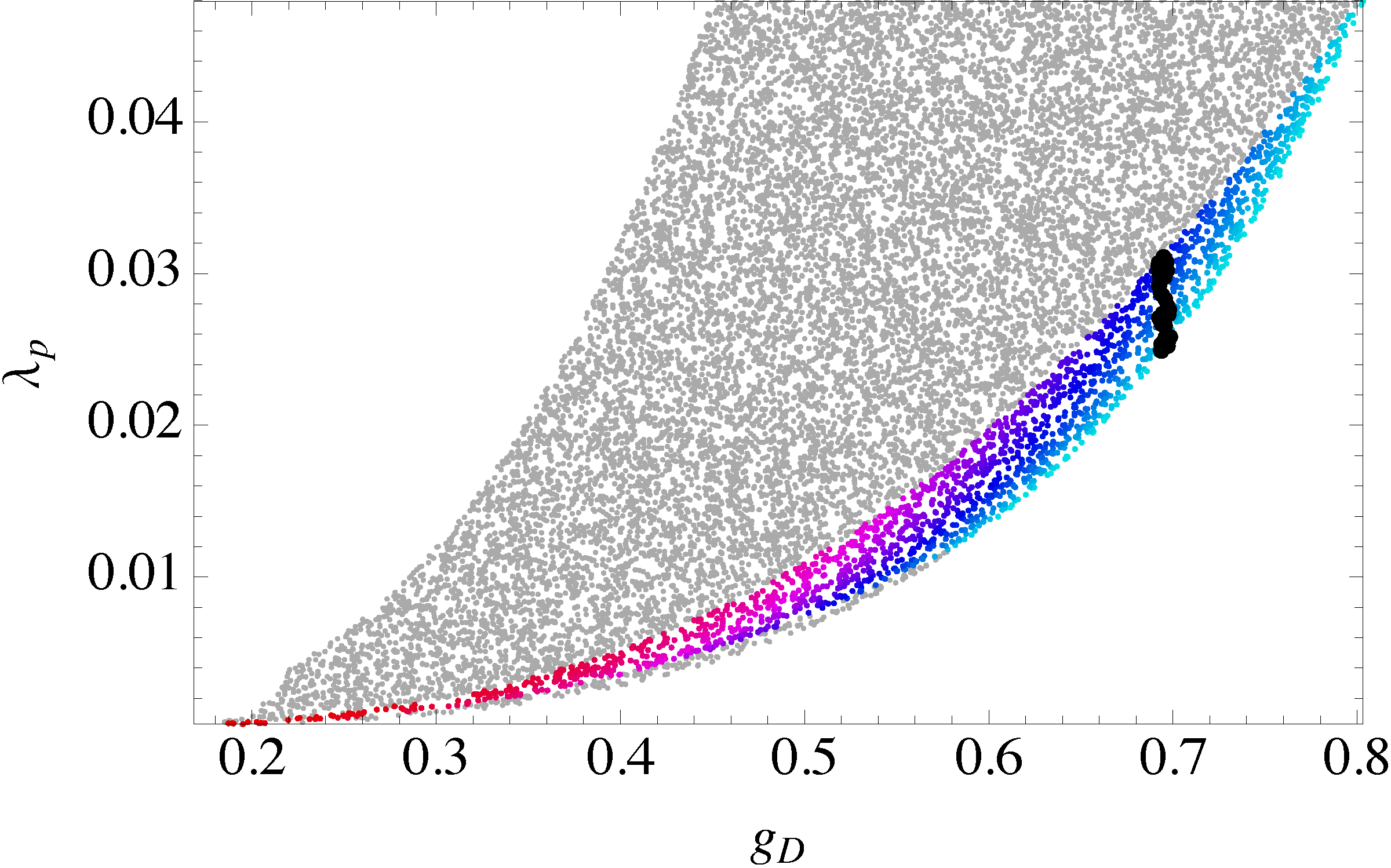}\hspace{0.1cm}
\caption{Portal coupling as a function of the dark gauge coupling for $N=2$ (left panel) and $N=3$ (right panel) in color for stable and perturbative data points, with color code function of $c_\alpha=\cos\alpha$ as given by the bar in the left panel. The data points that also produce an experimentally viable dark matter abundance are shown in black, and the gray points represent the unstable and/or non-perturbative data points which though feature a viable coupling coefficient $c_\alpha$.}
\label{gDlp}
\end{figure}

The SM potential turns out to be metastable for the measured Higgs mass, since the quartic Higgs self coupling turns to negative values at a scale smaller or equal to the Planck scale. The beta function of this coupling, Eq.~\eqref{blh}, receives in the present model an extra contribution from the portal coupling, which though small, is in principle enough to keep $\lambda_h$ positive up to the Planck scale \cite{Hambye:2013dgv}. Moreover the mixing of the scalar fields allows for larger values of $\lambda_h$ at the EW scale. Given that the beta function of $\lambda_\phi$,  Eq.~\eqref{bglD}, is numerically positive for viable data points, the value of $\lambda_\phi$, greater than zero at the EW scale because of Eq.~\eqref{minc}, stays positive at all scales. Only for $5\%$ of the roughly $4\times 10^4$ viable data points, selected in the previous section for each value of $N=2,3,4$, all couplings stay positive and perturbative (i.e. smaller than $2\pi$) up to the Planck scale:\be\label{pstabc}
0<\lambda_h,\lambda_\phi,g_D,y_t<2\pi \quad {\rm at}\quad v_w<\Lambda<M_{\rm{Pl}}\ .
\ee
The average values of the only two free parameters, $\lambda_p$ and $g_D$, with their respective standard deviations for the viable data points featuring perturbativity and stability are
\be
N=\left\{\begin{array}{c}
2\\ 3\\ 4
\end{array}\right.
\ , \quad
\lambda_p=\begin{array}{c}
0.020\pm0.011\\ 0.019\pm 0.011\\ 0.019\pm 0.010
\end{array}\ ,\quad 
g_D=\begin{array}{c}
0.55\pm 0.11\\ 0.60\pm 0.12\\ 0.63\pm 0.12
\end{array}\ .
\ee	
In Fig.~\ref{gDlp} we plot the portal coupling as a function of the dark gauge coupling for $N=2$ (left panel) and $N=3$ (right panel).  Stable and perturbative data points are in color, with color code function of $c_\alpha=\cos\alpha$ as given by the bar in the left panel. The black points represent the data points that also produce an experimentally viable dark matter abundance, as determined in the next subsection. Finally, the gray points represent the unstable and/or non-perturbative data points which though feature a viable coupling coefficient $c_\alpha$. 

\begin{figure}[htb]
\centering
\includegraphics[width=0.46\textwidth]{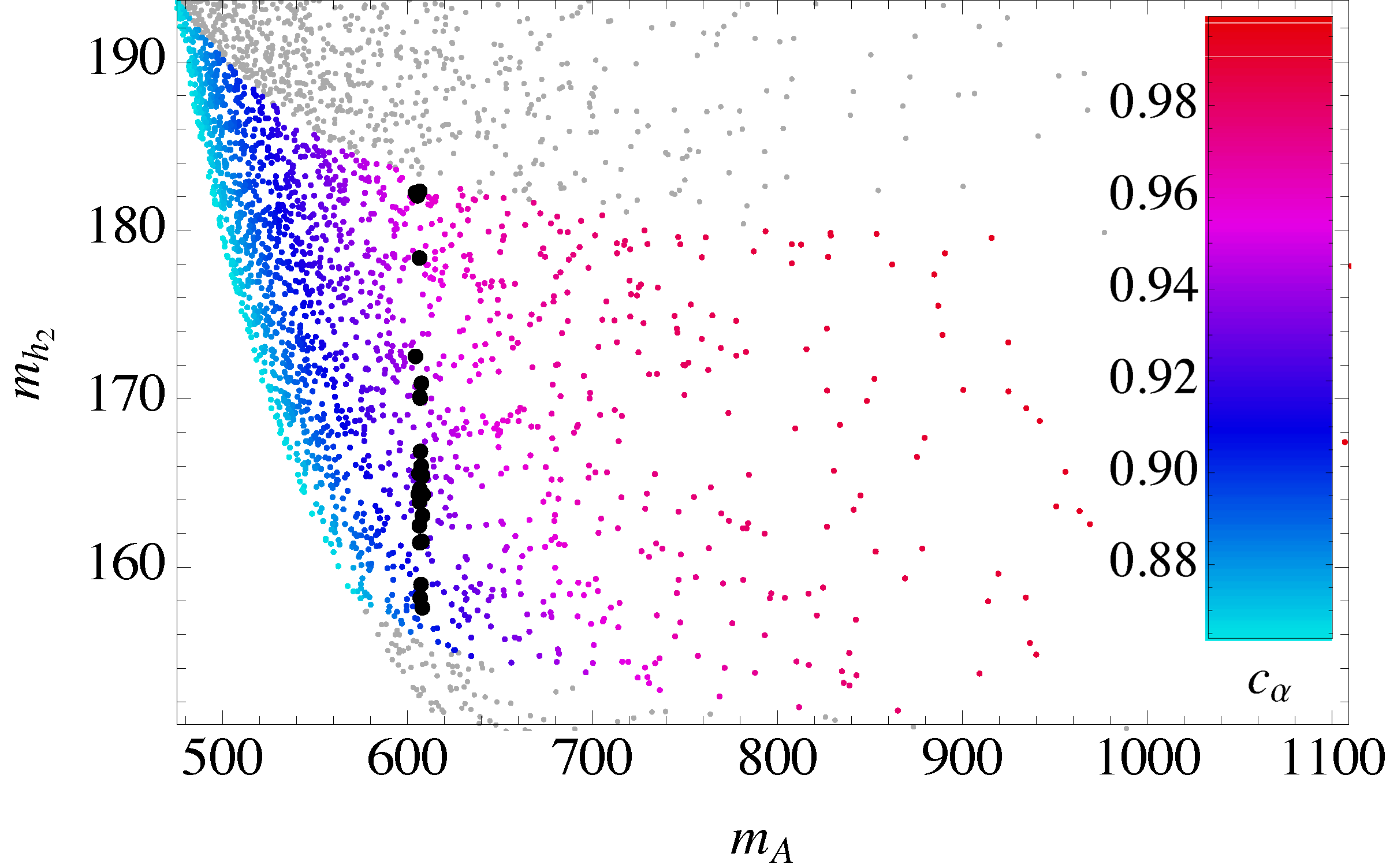}\hspace{1cm}
\includegraphics[width=0.46\textwidth]{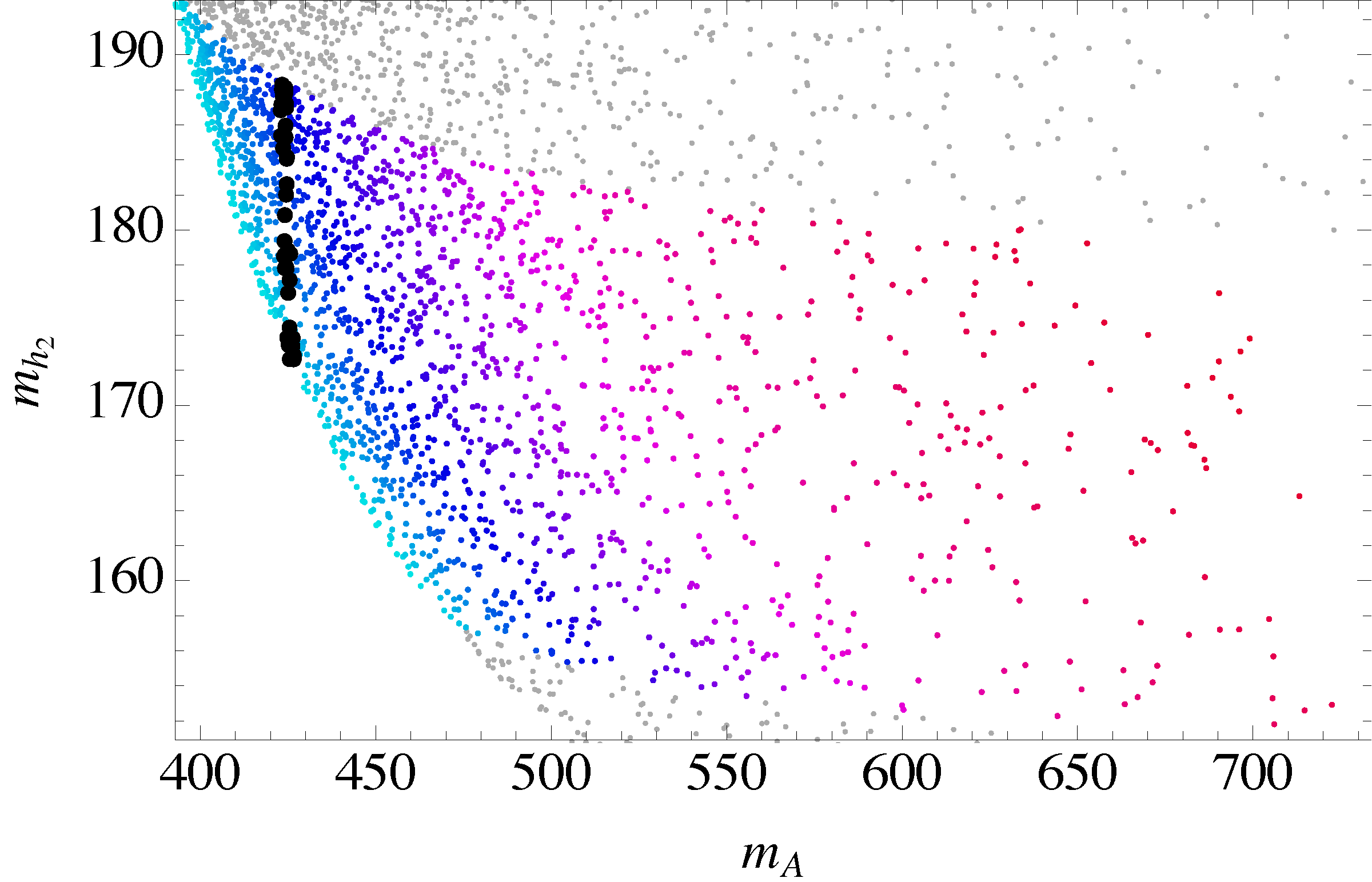}\hspace{0.1cm}
\caption{Dark scalar and dark vector masses for all the data points satisfying the LHC constraint, for $N=2$ (left panel) and $N=3$ (right panel), in color for those that also feature stability and perturbativity, in black those that satisfy the dark matter abundance constraint as well, while the data points that do not satisfy neither of the last two constraints are in gray.}
\label{mDmp}
\end{figure}

From Fig.~\ref{gDlp} one can see that for data points in color there is a strong correlation among $g_D$, $\lambda_p$, and $c_\alpha$, with $g_D$ constrained to larger values and $\lambda_p$ to smaller ones by the perturbativity and stability requirements. As a result of this correlation the allowed values of the mass of the dark Higgs $h_2$ fall in a very narrow range:
\be\label{mAh2}
N=\left\{\begin{array}{c}
2\\ 3\\ 4
\end{array}\right.
\ , \quad
m_{h_2}/{\rm GeV}=\begin{array}{c}
175\pm 10\\ 175\pm 10\\ 175\pm 9
\end{array}\ ,\quad 
m_A=\begin{array}{c}
580\pm 99\\ 480\pm 66\\ 420\pm 63
\end{array}\ .
\ee
In Fig.~\ref{mDmp} we plot the values of the dark scalar and dark vector masses for all the data points satisfying the LHC constraint, for $N=2$ (left panel) and $N=3$ (right panel). The color coding is the same as in Fig. \ref{gDlp}.

From Fig.~\ref{mDmp} one can see that the dark gauge boson mass is clearly correlated with the mixing coefficient $c_\alpha$. 
Although the model makes a very definite prediction for the range of the viable masses of the heavy physical Higgs $h_2$, the couplings of $h_2$ to SM particles are very small making its discovery at present colliders likely impossible. 

As a final comment we point out that no data point featuring a dark Higgs lighter than 125 GeV satisfies all the collider, stability, and perturbativity constraints simultaneously. We therefore do not investigate further the possibility that $h_2$ might be the Higgs boson discovered at LHC.

In the next section we implement in our analysis the dark matter abundance, as determined for the present model in Subsection~\eqref{DMs}, and direct detection constraints to further test the model's phenomenological viability.


\subsection{Dark matter abundance and direct detection}
\begin{figure}[htb]
\centering
\includegraphics[width=0.46\textwidth]{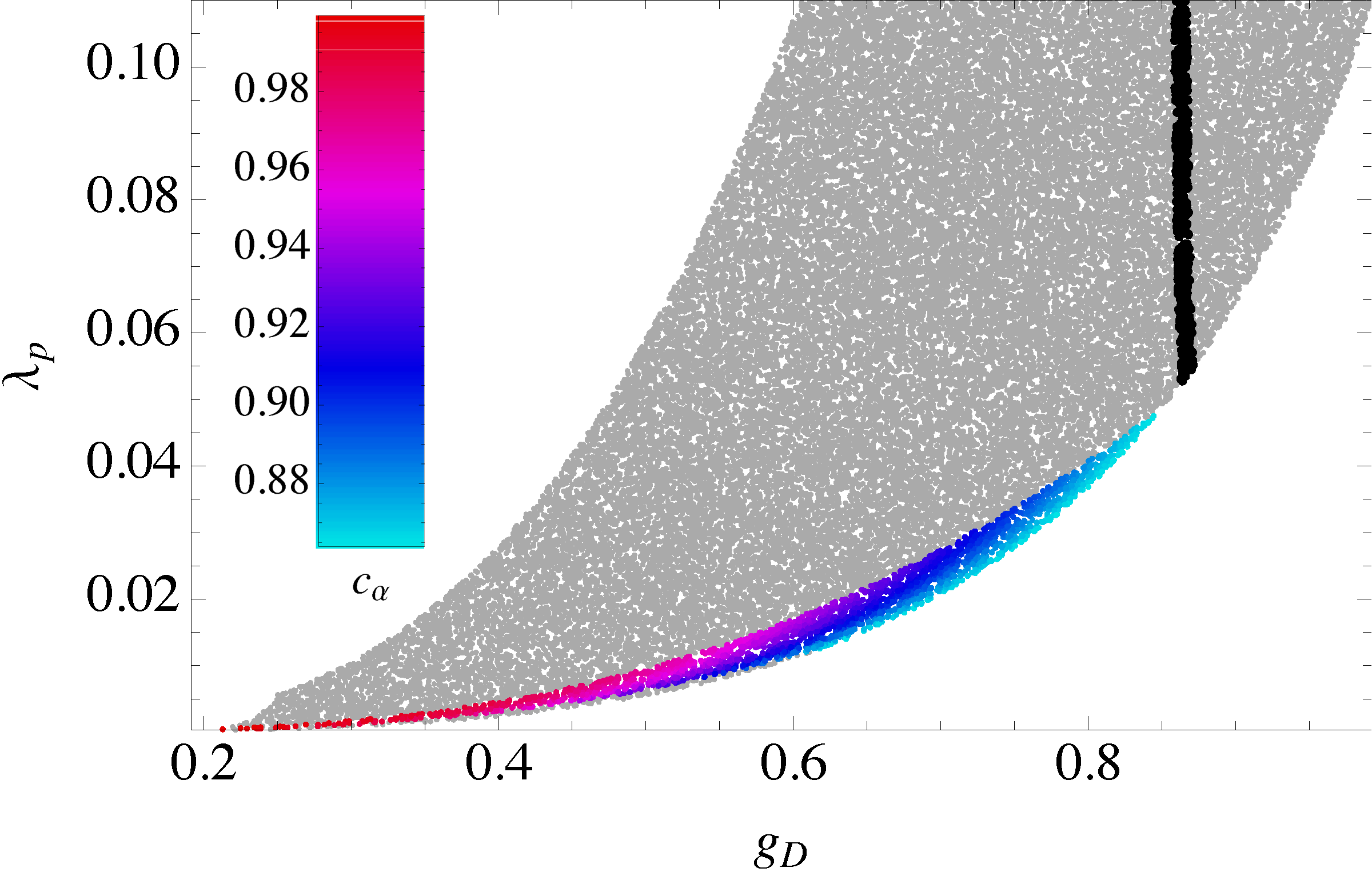}\hspace{1cm}
\includegraphics[width=0.46\textwidth]{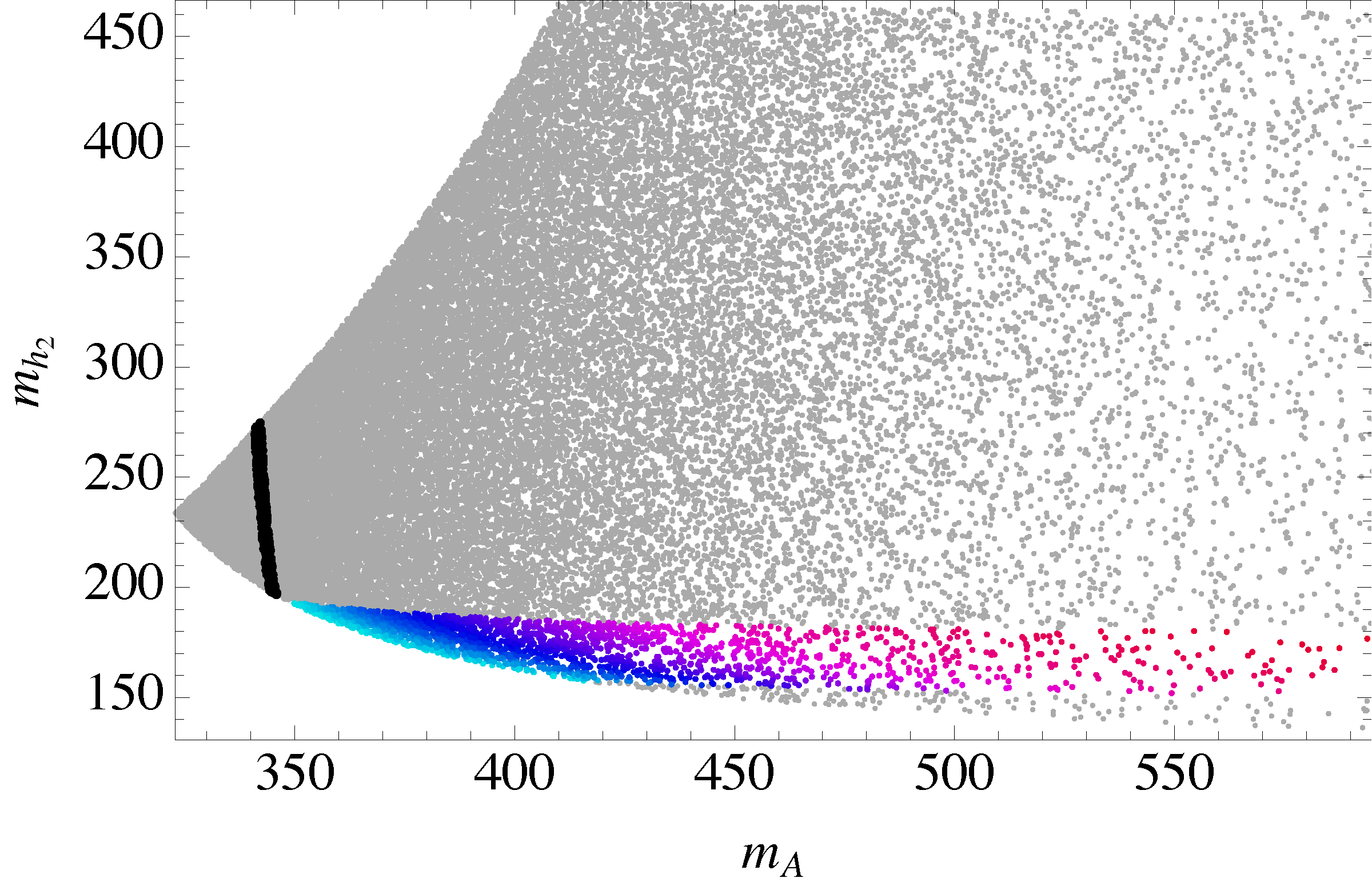}\hspace{0.1cm}
\caption{Portal coupling vs dark gauge coupling (left panel) and dark scalar vs dark gauge masses (right panel) for $N=4$, in color for stable and perturbative data points, with a color code function of $c_\alpha$ as given by the bar in the left panel, in black for data points that instead produce an experimentally viable dark matter abundance, and in gray for data points which satisfy only the LHC constraint on $c_\alpha$.}
\label{glmm}
\end{figure}

We compare the numerical result produced by Eq.~\eqref{Ohs} at each viable data point with the 95\% experimental interval \cite{Ade:2015xua}
\be\label{DMrd}
\Omega h^2=0.1193\pm0.0028\ ,
\ee
and find 23 data points (or 1\% of the total) for $N=2$ (Figs.~\ref{gDlp},\ref{mDmp}, left panels) and 39 (or 2\% of the total) for $N=3$ (Figs.~\ref{gDlp},\ref{mDmp}, right panels) that satisfy the dark matter constraint above as well as those in Eqs.~(\ref{LHCc},\ref{pstabc}). Interestingly, as can be seen from Fig.~\ref{gDlp}, for increasing $N$ the dark matter constraints and the constraints from stability pull in different directions in the $(g_D,\lambda_p)$-plane. Consequently, already for $N=4$ none of the data points satisfy all constraints, as can be seen from Fig.~\ref{glmm}.

Finally, we also impose the direct detection constraints. The spin independent cross section for the elastic scattering off a nucleon  $\cal N$ of the vector dark matter candidate, $A^a$, mediated by either $h_1$ or $h_2$ is, in the limit $m_A\gg m_{\cal N}$,
\be\label{sSI}
\sigma_{SI}\left({\cal N} A\rightarrow {\cal N} A\right)=\frac{f_{\cal N}^2 m_{\cal N}^4 m_A^2}{64 \pi  v_h^2 v_{\phi }^2} \sin^2 2 \alpha \left(\frac{1}{m_1^2}-\frac{1}{m_2^2}\right)^2\ ,
\ee
where $f_{\cal N}=0.303$ is the effective Higgs nucleon coupling \cite{Cline:2013gha}\footnote{We thank K. Kainulainen for discussions on this and for providing an updated value}, $m_{\cal N}=0.939$ GeV is the average nucleon mass. 
In the mass range of interest to us here, $m_D\sim{\cal O}(100)$ GeV, the most stringent bounds are provided by the LUX experiment \cite{Akerib:2013tjd}. We evaluate Eq.~\eqref{sSI} at each data point and compare with the experimental constraint in \cite{Akerib:2013tjd}, as a function of the mass of the dark matter candidate, $m_A$: all the data points satisfying the LHC constraint on the Higgs couplings, Eq.~\eqref{LHCc}, and consequentially also the points satisfying the stability, perturbativity, and dark matter abundance constraints, are viable. In more detail we obtain that the experimental constraint on the spin independent cross section is on average an order of magnitude larger than the predicted value
\be
N=\left\{\begin{array}{c}
2\\ 3
\end{array}\right.
\ , \quad
\sigma_{SI}\left({\cal N} A\rightarrow {\cal N} A\right)=\begin{array}{c}
(5\pm 4)\times 10^{-46}~ {\rm cm^2}\\ (3\pm 3)\times 10^{-46}~ {\rm cm^2}
\end{array}\ .
\ee

To summarize the results of this section, the classically conformal SU$(3)$ bi-adjoint scalar extended SM turns out to be an even more appealing model than its SU$(2)$ version, given that the former features a larger region of parameter space that satisfies collider, stability, perturbativity, and dark matter abundance constraints than the latter.

Vector DM in SU$(3)$ gauge theory has been considered also in \cite{Gross:2015cwa}. There SU$(3)$ is broken completely by a pair of scalar triplets, which introduce four new physical scalars, compared to just one in the present minimal model.


\section{Discoverability at LHC Run II}\label{LHCtest}

\begin{figure}[htb]
\centering
\includegraphics[width=0.46\textwidth]{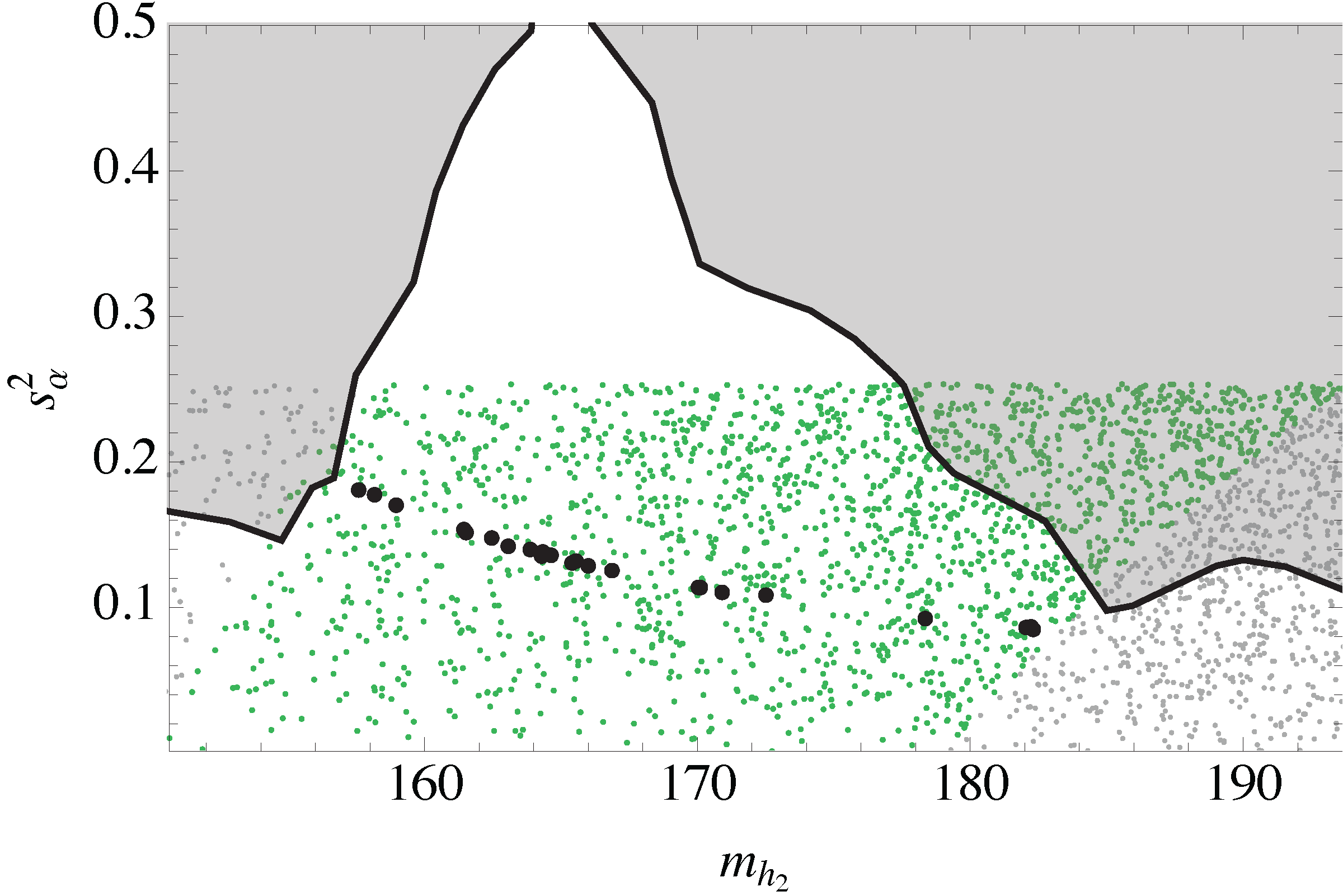}\hspace{1cm}
\includegraphics[width=0.46\textwidth]{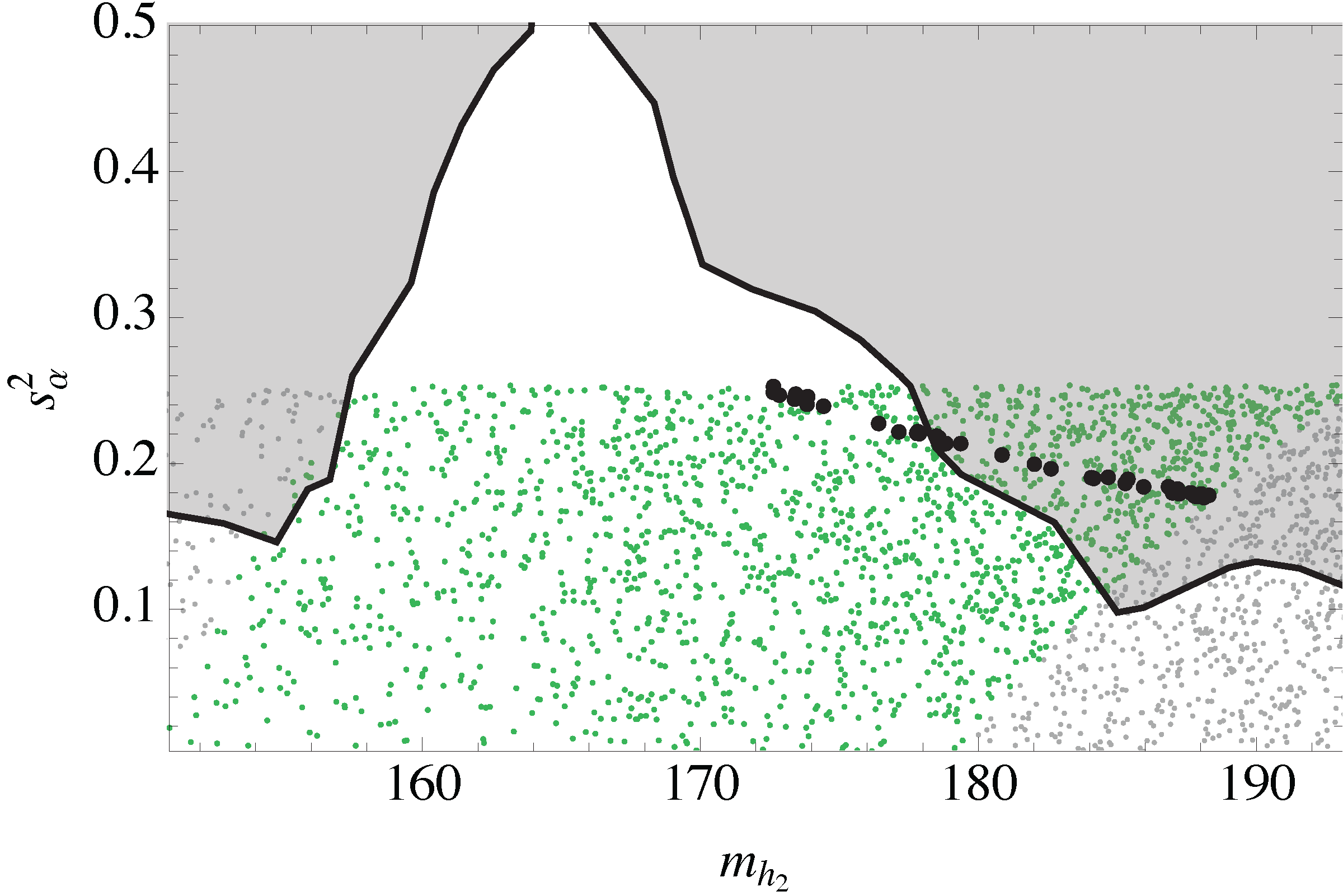}\hspace{0.1cm}
\caption{CMS constraint (shaded region ruled out at 95\%CL) on $s^2_\alpha=\sin^2\alpha$ in function of the heavy Higgs mass, together with the viable data points (in green those stable and with viable light Higgs couplings, and in black those that satisfy also DM constraints), for $N=2$ (left panel) and $N=3$ (right panel).}
\label{mh2s2}
\end{figure}

The bounds on additional Higgs-like resonances in the mass region defined by Eq.~\eqref{mAh2} are rather stringent \cite{Khachatryan:2015cwa}, given that such a heavy Higgs can decay into a pair of EW bosons almost at rest. The amplitude for production and decay of the heavy Higgs $h_2$ is equal to that of the SM Higgs suppressed by a factor $s^2_\alpha=\sin^2\alpha$, and there are no hidden decays to new particles or to a pair of light Higgs bosons. In Fig.~\ref{mh2s2} we plot the CMS constraint on $s^2_\alpha$ (Fig.~8 in \cite{Khachatryan:2015cwa}), together with the viable data points (in green those stable and with viable light Higgs couplings, and in black those that satisfy also DM constraints): of the universally viable data points, about half for $N=3$ (right panel) and all for $N=2$ (left panel) satisfy the CMS constraint. Notice that even assuming that not all the DM relic density is generated by the dark vectors $A^a$, only in the region above the strip of black data points the corresponding relic density satisfies the 95\% CL upper bound in Eq.~\eqref{DMrd}, while the region below is ruled out. 
To estimate the improvement of this upper bound at LHC Run II, we assume the corresponding constraint on the cross section to be dominated by data statistical uncertainty, and therefore to depend on the square root of the total number N of events observed:
\be\label{NLHC}
\sqrt{\rm N}=\sqrt{\sigma_{h_2}\epsilon_{eff} L_{tot}}
\ee
where $\sigma_{h_2}$ is the production rate of a SM-like Higgs boson of mass $m_{h_2}$, $\epsilon_{eff}$ is the efficiency of the trigger, and $L_{tot}$ the total integrated luminosity. Assuming the efficiency at Run II to be unchanged, and taking the total integrated luminosity at the end of Run II in 2019 to be 150~$fb^{-1}$, the upper bound on the production rate of $h_2$ should be reduced by a factor of
\be
\left(\sqrt{2.497\times \frac{150}{25}+1}\right)^{-1}=1/4\ ,
\ee
where the first coefficient under square root is equal to the ratio of production rates for a 175~GeV SM-like Higgs at 13 and 8~TeV, respectively. This result corresponds to a reduction of 1/2 of the upper limit on $s^2_\alpha$, which changes like the $h_2$ production amplitude. Assuming the limits in Fig.~\ref{mh2s2} to be simply shifted down by a factor of 1/2, we expect a large portion of viable parameter space of the SU$(2)$ model to be tested at LHC Run II, and a heavy Higgs to be discovered by the end of Run II or the SU$(3)$ model to be ruled out.


\section{Conclusions}\label{checkout}
In this paper we presented a novel extension of the SM, featuring a new scalar $\Phi$ in the bi-adjoint representation of SU$(N)_L\times$SU$(N)_R$, with only SU$(N)_L$ gauged. The vev of such $N^2-1$ dimensional matrix field is proportional to the identity matrix, and breaks completely the new gauge group, providing the corresponding vector bosons $A^a$ with the same mass.
Because of the residual SO$(N)$ global symmetry, $A^a$ is stable and a viable dark matter candidate. The dark sector couples to the SM only via a Higgs portal term. We set the mass parameters to zero, and let the EW symmetry be broken via dimensional transmutation due to quantum corrections. This choice has two motivations: 1) Reducing the number of free parameters to just two (the gauge coupling $g_D$ and the portal coupling $\lambda_p$), and 2) Allowing only for logarithmic quantum corrections to the scalar mass, and as a consequence solving in principle the SM hierarchy problem. In this extension of the SM indeed the fine tuning problem is traded with that of finding an ultraviolet boundary condition that motivates the choice of zero mass parameters. 

The resulting model provides a general setup for SU$(N)$ vector dark matter with a minimal number of free parameters  and matter fields. We studied quantitatively the phenomenology of the model for $N=2,3,4$ by scanning the two-dimensional parameter space for data points producing a viable mass spectrum. We then selected, by performing a goodness of fit analysis, the data points that match at 95\% CL the LHC measured coupling coefficients of the physical Higgs to SM particles. We then calculated the beta functions for all the couplings and showed that for about 5\% of the LHC viable data points the Higgs field quartic coupling (as well as the dark scalar one) stay positive up to the Planck scale, therefore solving the metastability problem of the SM. The same data points feature also successful EW symmetry breaking and perturbativity up to the Planck scale. Finally, we calculated the dark matter relic abundance and selected the data points that satisfy the experimental 95\% CL bound: for $N=2,3$ we found that about 1\% and 2\%, respectively, of the LHC viable, stable data points produce also a viable relic abundance, while for $N=4$ no such data point exists. The constraint from the dark matter direct detection experiments is comfortably satisfied by the same viable data points: they produce a cross section for the scattering of the dark matter candidate on nuclei which is an order of magnitude smaller than the experimental lower limit. To assess the discoverability of the predicted heavy Higgs, we imposed the CMS constraint on additional Higgs-like resonances and found that about half of the $N=3$ universally viable data points are actually already ruled out, while none is for the $N=2$ model.

To summarize, we have shown that the minimal vector dark matter extension of the SM presented in this paper leads to EW symmetry breaking through radiative corrections, stabilizes the scalar potential while providing an experimentally viable dark matter candidate and satisfying direct search and LHC Higgs coupling constraints. All  this is achieved within a two dimensional parameter space. As such, these models represent a valid and attractive scenario beyond the SM. We find that $N=2,3$ provide the most appealing model setting in light of present data from colliders and dark matter direct search experiments. We expect a heavy Higgs to be discovered at LHC by the end of Run II or the $N=3$ model to be ruled out.


\section*{Acknowledgements} 
We thank Alexander Belyaev, Kimmo Kainulainen, Venus Keus, and Lauri Wendland for helpful discussions and comments. We acknowledge the financial support from the Academy of Finland, project number 267842.

\newpage

\appendix
\section{Scalar Mass Matrix}
\label{mass}
The elements of the scalar mass matrix at one loop in the $(h,\sigma)$ basis are
\bea\label{Ms1L}
\left({\cal M}^2_\varphi\right)_{1 1}&=&\lambda _p v_{\phi }^2+\frac{1}{32\pi^2}\left\{\log \left[\left(1+\frac{v_{\phi }^2}{v_h^2}\right) \lambda _p\right] \left(v_h^2+\frac{9 v_{\phi }^4}{v_h^2}\right) \lambda _p^2-6 \lambda _p^2 v_{\phi }^2
+16 \left[1+3 \log \left(\frac{m_W}{v_h}\right)\right] \frac{m_W^4}{v_h^2}+\right.\nonumber\\
& &\quad\quad\quad\quad\quad\quad\quad\left.8 \left[1+3 \log \left(\frac{m_Z}{v_h}\right)\right] \frac{m_Z^4}{v_h^2}-96 \log \left(\frac{m_b}{v_h}\right) \frac{m_b^4}{v_h^2}-96 \log \left(\frac{m_t}{v_h}\right) \frac{m_t^4}{v_h^2}\right\},\nonumber\\
\left({\cal M}^2_\varphi\right)_{1 2}&=&\left({\cal M}^2_\varphi\right)_{2 1}=- \lambda _p v_h v_{\phi }+\frac{1}{32\pi^2}\left\{6 \lambda _p v_h v_{\phi } -\log \left[\left(1+\frac{v_{\phi }^2}{v_h^2}\right) \lambda _p\right] \left(\frac{3 v_h^3}{v_{\phi }}-4 v_{\phi } v_h+\frac{3 v_{\phi }^3}{v_h}\right)
   \lambda _p^2\right\},\nonumber\\
\left({\cal M}^2_\varphi\right)_{2 2}&=&\lambda _p v_{h }^2+\frac{1}{32\pi^2}\left\{\log \left[\left(1+\frac{v_{\phi }^2}{v_h^2}\right) \lambda _p\right] \left(v_{\phi }^2+\frac{9 v_h^4}{v_{\phi }^2}\right) \lambda _p^2-6 \lambda _p^2  v_h^2+\right.\nonumber\\
& &\quad\quad\quad\left.8\left(N^2 -1\right)\left[1+3 \log \left(\frac{m_A}{v_h}\right)\right] \frac{m_A^4}{v_{\phi }^2}\right\},
\label{scalarmasses}
\eea
with the SM masses given by the usual expressions , $m_A$ by Eq.~\eqref{mA}, and the renormalization scale set equal to $v_h$.


\section{Beta Functions}\label{apbeta}
The only SM beta function that is modified in the present model is
\be\label{blh}
16 \pi^2\frac{d \lambda_h}{d t}=16 \pi^2\left(\frac{d \lambda_h}{d t}\right)_{SM}+N^2 \lambda_p^2 ,
\ee
with $t=\log(E/\Lambda)$. The beta functions for the beyond the SM couplings in Eqs.~(\ref{Vt},\ref{LP}) are, for $N=2,3,4$:
\bea\label{bglD}
16 \pi^2\frac{d g_D}{d t}&=&-r_{1,N} g_D^3\ ;\quad r_{1,N}=\frac{253}{36},\frac{43}{4},\frac{1297}{90}\ ;\nonumber\\
16 \pi^2\frac{d \lambda_\phi}{d t}&=&r_{2,N} g_D^4-r_{3,N} g_D^2 \lambda _{\phi }+r_{4,N} \lambda _{\phi }^2+4 \lambda _{p }^2\ ;\quad r_{2,N}=\frac{41}{6},\frac{51}{16},\frac{353}{150}\ ;\quad r_{3,N}=14,\frac{27}{2},\frac{76}{5}\ ;\nonumber\\
\ r_{4,N}&=&12,17,24\ ;\nonumber\\
16 \pi^2\frac{d \lambda_p}{d t}&=& -\frac{9}{10} g_1^2 \lambda _p-\frac{9}{2} g_2^2 \lambda _p-\frac{r_{3,N}}{2} g_D^2 \lambda _p+6 y_b^2 \lambda _p+6 y_t^2 \lambda _p+2 y_{\tau }^2 \lambda _p+6 \lambda _h \lambda _p-4 \lambda
   _p^2+r_{5,N} \lambda _p \lambda _{\phi }\ ;\nonumber\\ r_{5,N}&=&6,11,18 \ .
\eea



\begin{thebibliography}{99}
\bibitem{CMS} 
  CMS Collaboration [CMS Collaboration],
and studies of the compatibility of its couplings with the standard model,''
  CMS-PAS-HIG-14-009.
\end{thebibliography}
\end{document}